\documentclass[singlespacing,twocolumn]{elsart5p}
\usepackage{ae}
\usepackage{amsmath}
\usepackage{amssymb}
\usepackage{graphics}
\usepackage{graphicx}
\usepackage{epsfig}
\usepackage{dcolumn}
\usepackage{bm}

\begin{document}
\begin{frontmatter}

\title{Development of linseed oil-free Bakelite Resistive Plate Chambers}
\author[label1]{S. Biswas},
\author{S. Bhattacharya\thanksref{label2}\corauthref{cor1}},
\ead{sudeb.bhattacharya@saha.ac.in}
\thanks[label2]{}
\corauth[cor1]{}
\author[label2]{S. Bose},
\author[label1]{S. Chattopadhyay},
\author[label2]{S. Saha},
\author[label3]{Y.P. Viyogi}

\address[label1]{Variable Energy Cyclotron Centre, 1/AF Bidhan
Nagar, Kolkata-700 064, India}
\address[label2]{Saha Institute of Nuclear Physics, 1/AF Bidhan Nagar, Kolkata-700
064, India}
\address[label3]{Institute of Physics, Sachivalaya Marg, Bhubaneswar, Orissa-751 005, India}

\begin{abstract}
In this paper we would like to present a few characteristics of the
Resistive Plate Chambers (RPC) made of a particular grade of
bakelite paper laminates (P-120, NEMA LI-1989 Grade XXX), produced
and commercially available in India. This particular grade is used
for high voltage insulation in humid conditions. The chambers are
tested with cosmic rays in the streamer mode using argon,
tetrafluroethane and isobutane in 34:59:7 mixing ratio. In the first
set of detectors made with such grade, a thin coating of silicone
fluid on the inner surfaces of the bakelite was found to be
necessary for operation of the detector. Those silicone coated RPCs
were found to give satisfactory performance with stable efficiency
of $>$ 90\% continuously for a long period as reported earlier.
Results of the crosstalk measurement of these silicone coated RPC
will be presented in this paper. Very recently RPCs made with the
same grade of bakelite but having better surface finish, are found
to give equivalent performance even without any coating inside.
Preliminary results of this type of RPCs are also being presented.

\end{abstract}
\end{frontmatter}

\section{Introduction}
\label{}

In the proposed India-based Neutrino Observatory (INO), the RPCs
\cite{RSRC81} have been chosen as the active detector for muon
detection in an Iron Calorimeter (ICAL) \cite{INO06}. As proposed
presently, the ICAL is a sampling calorimeter consisting of 140
layers of magnetized iron, each of 60 mm thickness, using RPCs of 2
m $\times$ 2 m area as active media sandwiched between them. A 50
kton ICAL is expected to consist of about 27000 RPC modules. For the
ICAL-RPCs, main design criteria are (a) moderate position resolution
($\sim$ 1 cm), (b) good timing resolution ($\sim$ 1-2 ns) (c) ease
of fabrication in large scale with modular structure and most
importantly (d) low cost. The detailed R\&D are being performed on
the glass RPCs for INO \cite{NKM08}. A parallel effort on building
and testing of the RPC modules using the bakelite obtained from the
local industries in India is also going on.

At the end of nineties, it was found that the RPCs based on
bakelite show serious ageing effects reducing the efficiency
drastically \cite{JV03}. Detailed investigations revealed that the
use of linseed oil for the surface treatment in such cases was the
main reason for this ageing effect \cite{FA03,FAN03}. Efforts were
subsequently made to look for alternatives to linseed oil treatment,
or to develop bakelite sheets which can be used without the
application of linseed oil \cite{JZ05}.

The aim of the present study is to achieve stable performance of the
bakelite RPC detector without linseed oil for prolonged operation.
The method of construction of these RPCs and  the results of the long
term test has been reported earlier \cite{SB08}. In this article we
report some other characteristics of the silicone coated bakelite
RPC and the initial results of the RPC fabricated without any
coating.

\section{Design of the prototype RPCs}
\label{}

The RPCs were made of two 300 mm $\times$ 300 mm $\times$ 2 mm
bakelite sheets, used as electrodes, separated by a 2 mm gas gap. A
uniform separation of the electrodes was ensured by using five
polycarbonate button spacers of 10 mm diameter and 2 mm thickness,
and edge spacers of 300 mm $\times$ 8 mm $\times$ 2 mm dimension.
Two polycarbonate made nozzles were used for gas inlet and outlet
\cite{SB08Ar, SB08Pr}.

The high voltages (HV) to the RPC were applied on the graphite
coating (surface resistivity $\sim$ 1 M$\Omega$/$\Box$) made over
the outer surfaces of the bakelite. The induced RPC signals were
collected using copper and foam based pick-up strips, each of area
300 mm $\times$ 25 mm with a separation of 2 mm between two adjacent
strips. The pick-up strips were covered with 100 $\mu$m thick kapton
foils to isolate them from the graphite layers.

Premixed gas of Argon, Isobutane and Tetrafluroethane (R-134a) was
used in 34:7:59 mixing ratio. A typical flow rate of 0.4 ml per
minute was maintained by the gas delivery system, \cite{SBose08Pr},  
resulting in $\sim$ 3 changes of gap volume per day.

\section{Cosmic ray test setup}
\label{}

The RPCs were tested in the same cosmic ray test bench described in
Ref. 6. The coincidence between the signals obtained from the
scintillator I (350 mm $\times$ 250 mm size), the finger
scintillator(III) (200 mm $\times$ 40 mm size) which was placed
above the RPC plane and the scintillator II (350 mm $\times$ 250 mm
size) which was placed below, was taken as the Master trigger.
Finally, the ORed signal obtained from two adjacent pick-up strips
of the chamber was put in coincidence with the master trigger
obtained above. This was referred to as the coincidence trigger of
the RPC. The efficiency of the RPC detector, taken as the ratio
between the coincidence trigger rates of the RPC and the master
trigger rates of the 3-element plastic scintillator telescope was
measured over an area 200 mm $\times$ 40 mm which was the window of
the cosmic ray telescope. The average master trigger rate was
$\simeq$ 0.005 Hz/cm$^2$.

\begin{figure}
\includegraphics[scale=0.47]{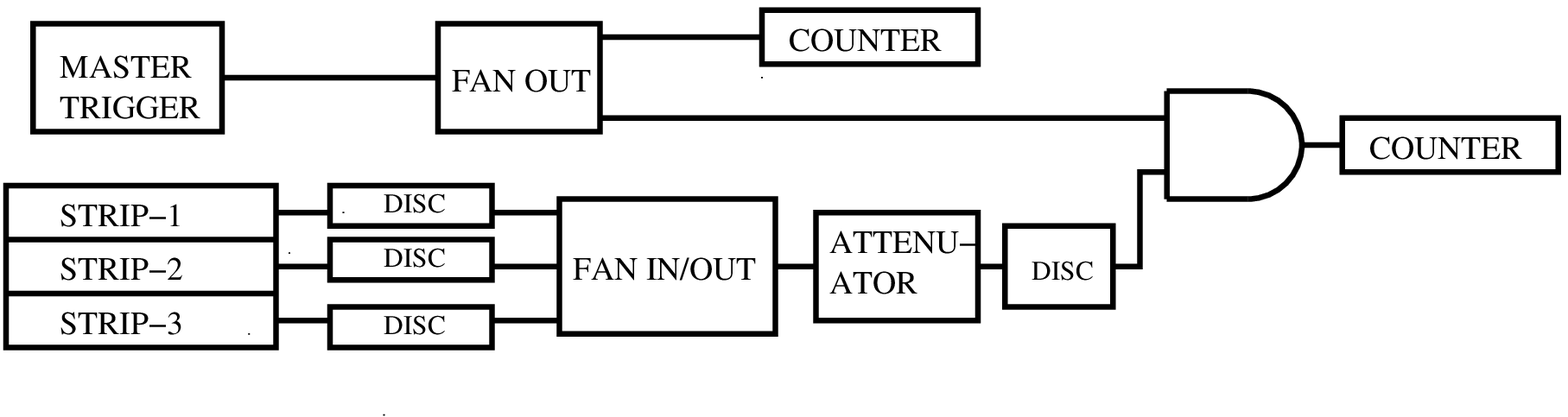}\\
\caption{\label{fig:epsart}Schematic representation of the crosstalk
measurement setup.} \label{fig:1}
\end{figure}

Fig. 1 shows the schematic of the crosstalk measurement setup. The
width of the finger scintillator covered one pick-up strip
completely and two adjacent strips partially. The signals from these
three strips after leading edge discriminator (LED threshold : 40
mV) were sent to the input of a fan in module. The output signal of
the previous stage after attenuation was again put to another
discriminator. The signal from this discriminator was taken in
coincidence with the master trigger. The crosstalk was defined by
the ratio of this coincidence count and the master trigger. The
attenuation factor was set at 0.3. The fan out signal after
attenuation, observed in the oscilloscope, was $\sim$ 250 mV when
one strip fires. The resulting pulse heights were $\sim$ 440 mV and
630 mV respectively, when signals from two strips and three strips
came simultaneously. In order to measure the crosstalk between two
and three adjacent strips, the discriminator threshold to the fan
out signal was set at 280 mV and 520 mV, respectively.

\section{Results}
\label{}

The following properties have been studied in the test setup for all
the chambers: (a) the efficiency of the chambers and their variation
with change in different parameters {\it e.g}. HV, gas composition,
laboratory environment etc., (b) the variation of the counting rate,
(c) the leakage current and their variation and (d) the long term
stability in the streamer mode. The long term behavior of the
silicone coated RPCs in the streamer mode has been reported earlier
\cite{SB08Pr}. Some more results are presented in this section.

The efficiency and the counting rate, which is also known as the
noise rate, of the RPCs have been studied by varying the applied HV
and setting different discriminator threshold values. The variation
of the efficiency and the counting rates with the applied HV for
three discriminator threshold value of a silicone coated RPC is
shown in Fig. 2. It is seen that the counting rate of the RPC
decreases with the increase of the threshold value which is expected
with the suppression of noise at a higher threshold value. However,
the efficiency curves do not depend much on the threshold setting
varying between 20 mV to 80 mV, except that the efficiency plateau
is marginally higher at the lowest threshold setting of 20 mV. A
typical screen dump of the oscilloscope pulse at 8 kV is shown in
Fig. 3. The figure shows the rise time is about 6 ns, which
indicates the fastness of the pulse.

\begin{figure}
\includegraphics[scale=0.4]{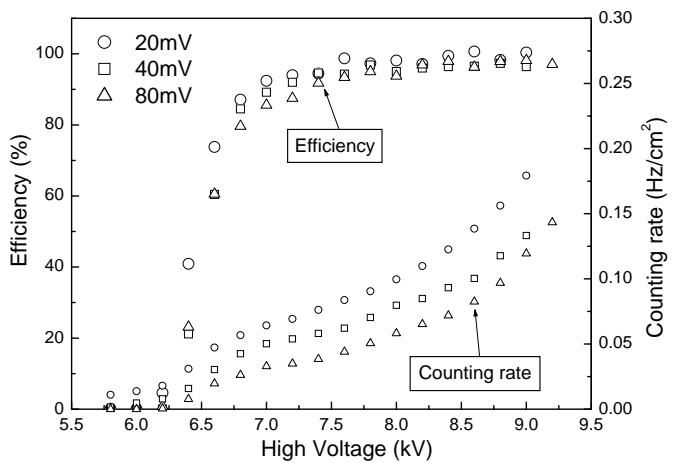}\\
\caption{\label{fig:epsart}The efficiency and the counting rate as a
function of high voltage for a silicone coated RPC made of P-120
grade bakelite with discriminator threshold values of 20 mV, 40 mV
and 80 mV. The gas mixture used was Argon (34\%) + Isobutane (7\%) +
R-134a (59\%). } \label{fig:2}
\end{figure}

\begin{figure}
\includegraphics[scale=0.37]{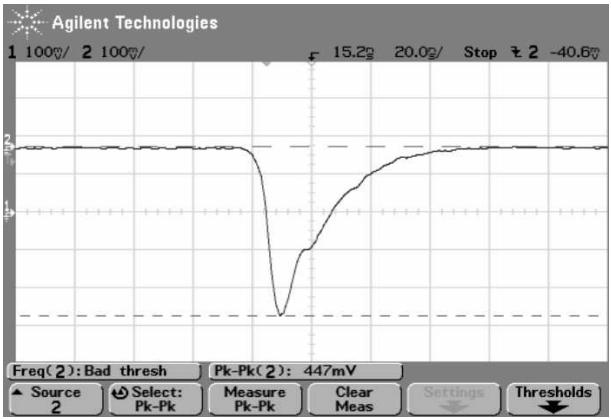}\\
\caption{\label{fig:epsart}Average induced pulse on a pick-up strip
at 8 kV (100 mV/Div, 20 ns/Div, 50 $\Omega$ load) of a silicone
coated P-120 grade bakelite RPC. } \label{fig:3}
\end{figure}

The effect of the environmental humidity on the efficiency curves
has also been studied. This measurement was done at relative
humidities of 58\% and 67\% of the laboratory environment and at the
room temperature of $\sim$ 22-25$^{\circ}$C. These curves, plotted
in Fig. 4, indicate no effect of humidity on the efficiency.
However, the counting rate (shown in Fig. 4) and the leakage
currents, measured simultaneously and plotted in Fig. 5, are larger
at higher humidity. These observations indicate that charge leakage
through the exterior surfaces may be contributing more at higher
humidity.

\begin{figure}
\includegraphics[scale=0.4]{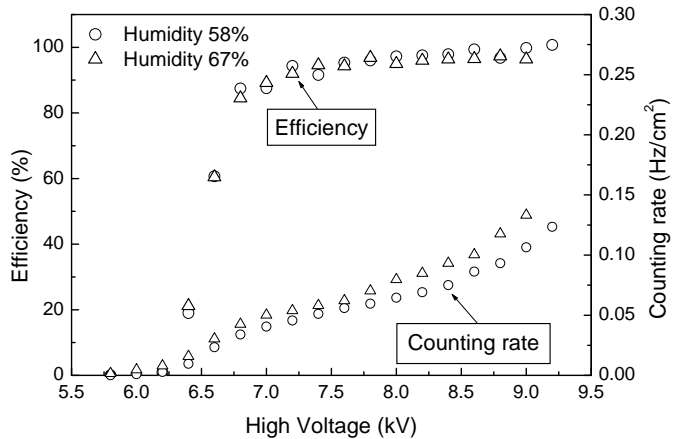}\\
\caption{\label{fig:epsart}The efficiency and the counting rate
versus high voltage for different humidities for a silicone coated
P-120 grade bakelite RPC.}\label{fig:4}
\end{figure}

\begin{figure}
\includegraphics[scale=0.4]{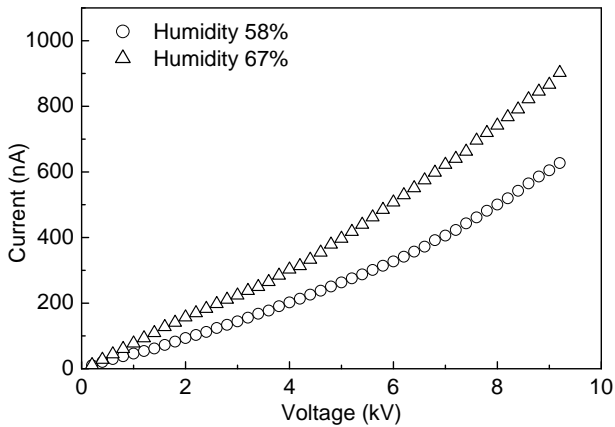}\\
\caption{\label{fig:epsart}Current versus high voltage for different
humidities for a silicone coated P-120 grade bakelite
RPC.}\label{fig:5}
\end{figure}

The long cable drive of RPC streamer pulse has been tested using
RG-174/U coaxial cables. A maximum of 40 m RG-174/U coaxial cable
has been used. The average pulse height of RPC in streamer mode was
$\sim$ 300-500 mV as shown in Fig. 3. The signal amplitude was
attenuated to $\sim$ 80\% after 40 m cable drive. The rise time
increases slightly at that length of cable. The variation of
normalised pulse height and rise time are shown in Fig. 6.

\begin{figure}
\includegraphics[scale=0.4]{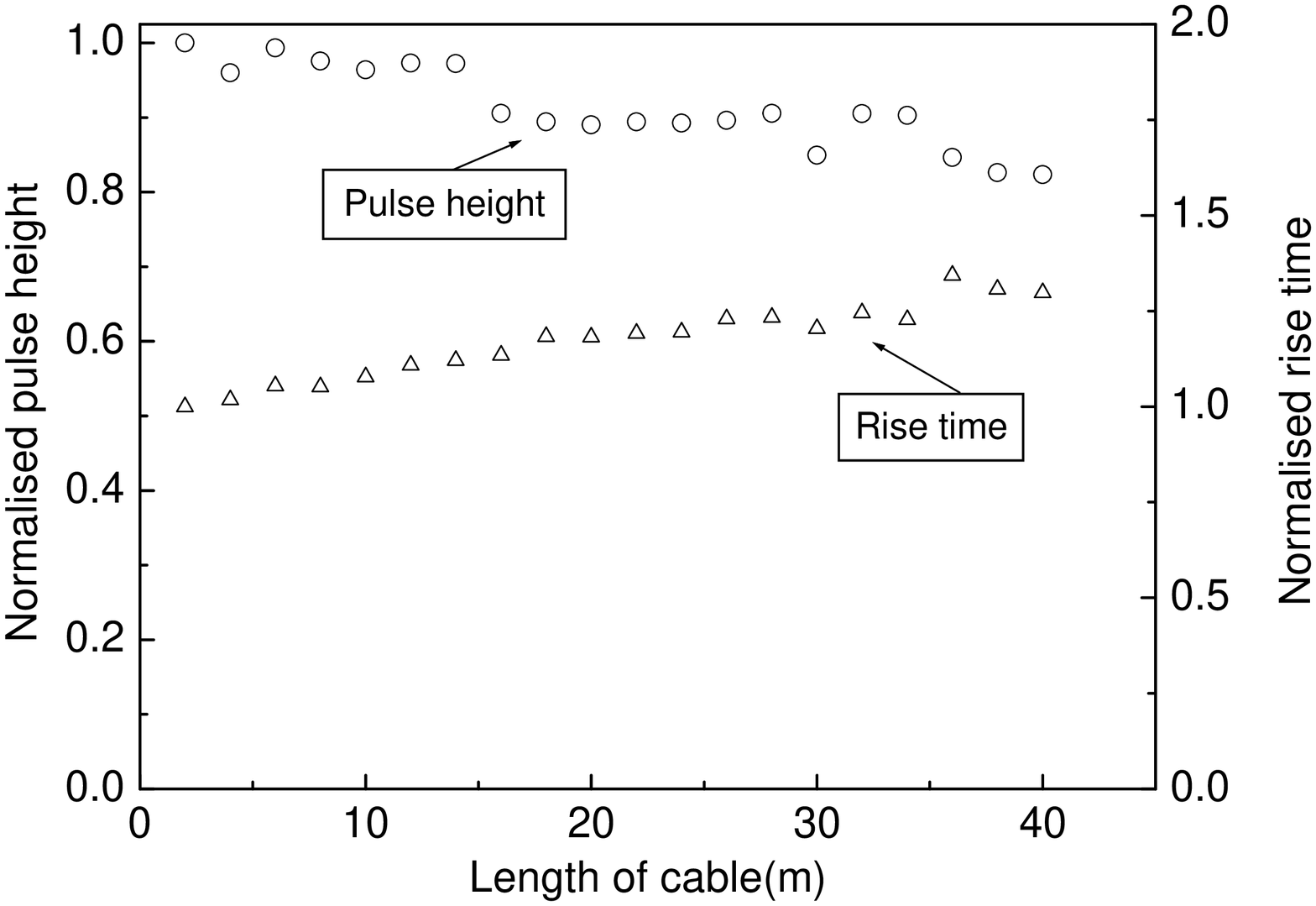}\\
\caption{\label{fig:epsart}Variation of the normalised pulse height
and rise time with the cable length for a silicone coated P-120
grade bakelite RPC.}\label{fig:6}
\end{figure}

The measurement of the crosstalk (CT) for a silicone coated RPC was
carried out rigorously. When a single particle induces signal on two
or more strips then the term crosstalk comes into the picture. The
crosstalk between the two and three RPC strips, defined in Section
3, had been measured by varying the applied HV and is shown in Fig
7. When the discrimination threshold after attenuation was set at
280 mV as stated in Section 3, the signals coming from the two
adjacent strips as well as three adjacent strips simultaneously
contributed to the crosstalk and it was found to be $<$ 20\% (Fig
7). The crosstalk between the three adjacent strips was found to be
$<$ 5\%. These values of the crosstalk have been taken into account
while estimating the final efficiency.

\begin{figure}
\includegraphics[scale=0.4]{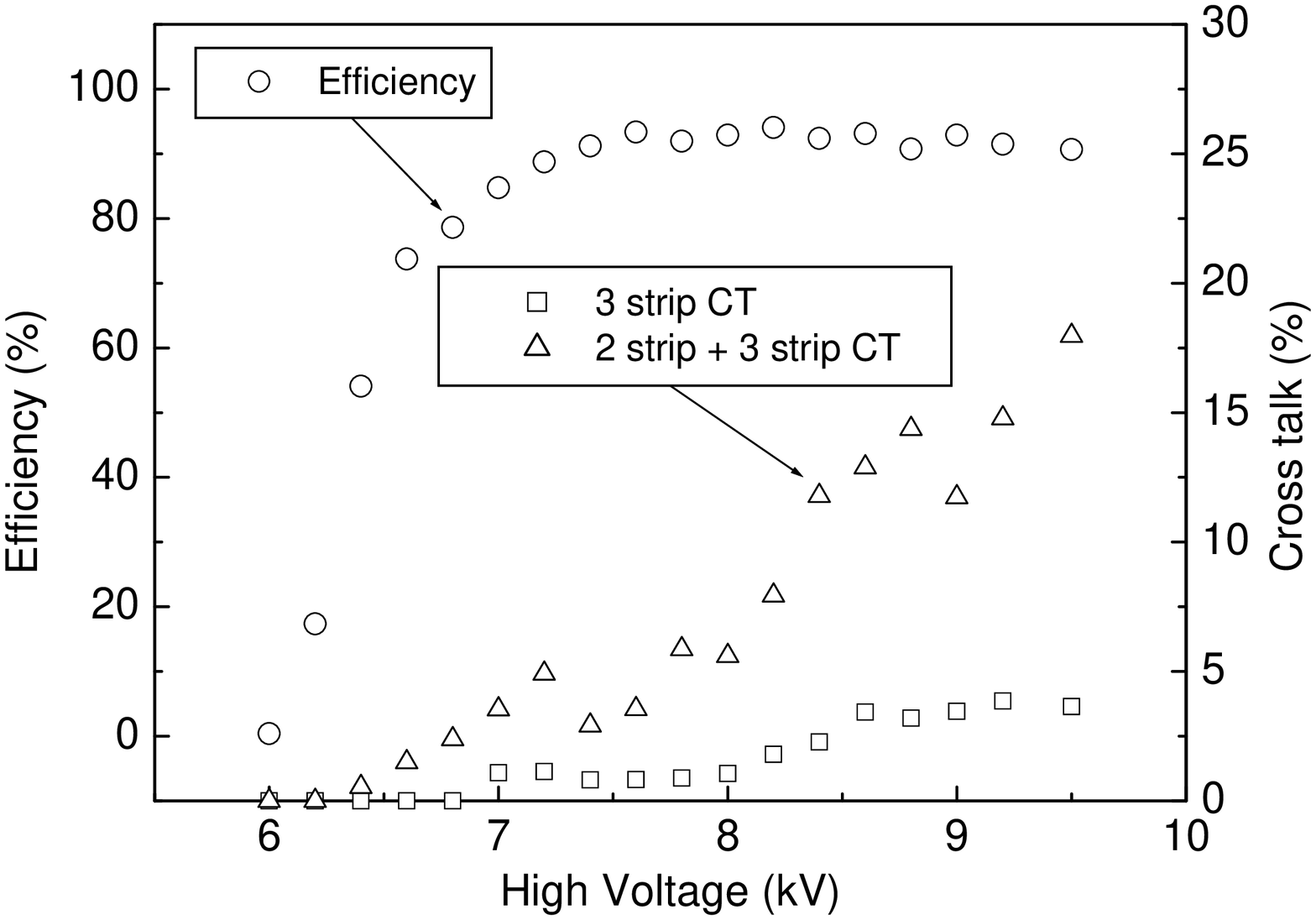}\\
\caption{\label{fig:epsart}Variation of the crosstalk with the high
voltage in comparison with the efficiency for a silicone coated
P-120 grade bakelite RPC.}\label{fig:7}
\end{figure}

The time resolution of a silicone coated RPC was measured and it was
found to be $\sim$ 2 ns (FWHM of the time spectra) at 8 kV. The
variation of time resolution with HV of that particular detector is
shown in Fig. 8.

\begin{figure}
\includegraphics[scale=0.4]{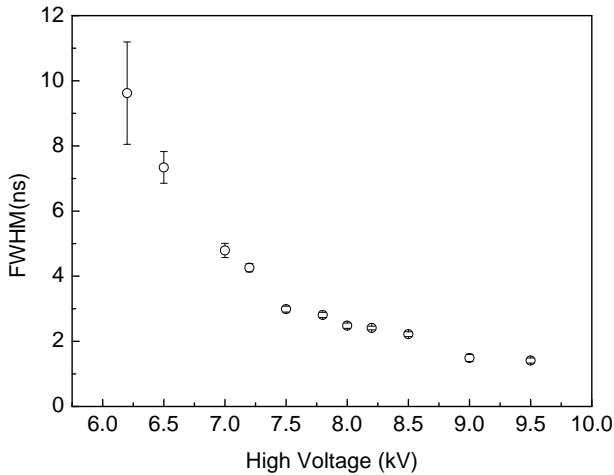}\\
\caption{\label{fig:epsart}The time resolution as a function of HV
for a silicone coated P-120 grade bakelite RPC.}\label{fig:8}
\end{figure}

Finally, one module was made with 1.6 mm thick P-120 grade bakelite
sheets with better surface finish. No coating was applied to that
RPC. The efficiency curve and the variation of the counting rate
with HV is plotted in Fig. 9. The current for this particular RPC
was found to be higher than that of the silicone coated one. It was
about 2 $\mu$A at 8 kV.

\begin{figure}
\includegraphics[scale=0.4]{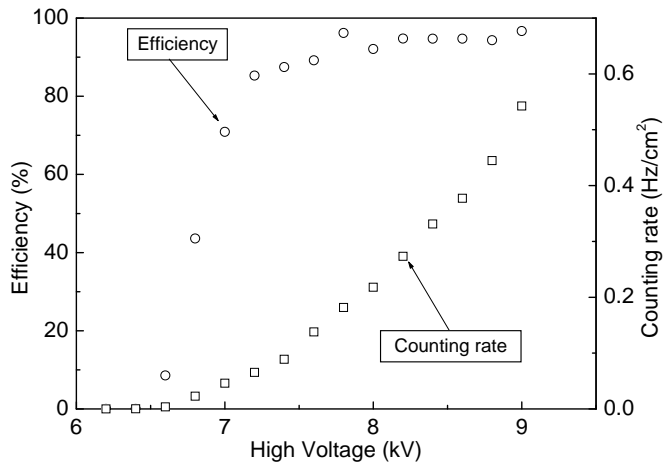}\\
\caption{\label{fig:epsart}The efficiency and the counting rate as a
function of HV for uncoated P-120 grade bakelite RPC.}\label{fig:9}
\end{figure}

\section{Conclusions and outlook}
\label{}

In conclusion, a rigorous study of RPCs made from a particular grade
of bakelite commercially available in India has been performed. An
efficiency of $>$ 90\% is obtained for silicone coated P-120 grade
bakelite RPC. The effect of threshold and external humidity on the
performance of the RPCs are presented in this paper. The current as
well as the counting rate increases with the increase of
environmental humidity.

The RPC streamer signals can be driven for a long distance without
any significant attenuation. The crosstalk between two neighboring
strips (which may be due to some real event) is found to be about
15\% and that between three neighboring strips is about 5\%. The
measured time resolution of a particular silicone coated RPC is
found to be $\sim$ 2 ns which is comparable to any single gap glass
or linseed oil coated bakelite RPC. Lastly the preliminary data of a
uncoated RPC is reported which show encouraging results, as one can
plan for large RPC (1 m $\times$ 1 m) with this particular grade of
bakelite with better surface finish and without any coating.

Further studies in this direction are being carried out which
include timing measurements of silicone coated RPCs and performance
of oil-less RPCs. These will be reported at a later stage.

\section{Acknowledgement}
\label{}

We are thankful to Prof. Naba Kumar Mondal of TIFR, India and Prof.
Kazuo Abe of KEK, Japan for their encouragement and many useful
suggestions in course of this work. We acknowledge the service
rendered by Mr. Ganesh Das of VECC for meticulously fabricating the
detectors. We would like to thank the SINP and VECC HEP workshop for
making the components of the detectors. Finally we acknowledge the
help received from the scientific staff of Electronics Workshop
Facility of SINP for building the gas flow control and delivery
system of the gas mixing unit used in this study.

\end{document}